# Investigation of multiple echo signals formation mechanism in magnet at excitation by two arbitrary radio-frequency pulses.

**M.D. Zviadadze*, G.I. Mamniashvili, R.L. Lepsveridze**,

Andronikashvili Institute of Physics, 6 Tamarashvili Str., Tbilisi, 0177, Georgia

A.M. Akhalkatsi,

Javakhishvili Tbilisi State University, 3 Chavchavadze Ave. Tbilisi, 0128, Georgia

*Corresponding author
E-mail address:  m.zviadadze@mail.ru,   m.zviadadze@aiphysics.ge

## Abstract

The quantum-mechanical calculations of intensities and time moments of appearance of multiple spin echo signals of excitation of nuclear spin system of magnet by two arbitrary width radio-frequency pulses were carried out. This method was used by us earlier at consideration of multiple-pulse analogs of single-pulse echo in multidomain magnets upon sudden jumps of the effecting magnetic field in the rotating coordinate system during the action of radio-frequency pulse.

The formation mechanisms of echo signals are discussed. The appearance of four primary stimulated echo signals is predicted. The total number of echo signals at fixed parameters of radio-frequency pulses does not exceed thirteen ones.

Theoretical conclusions are in compliance with experiments carried out on lithium ferrite. As it was established by us earlier in this magnetic dielectric, in difference from ferrometals, it is observed very short relaxation times of single-pulse and two-pulse stimulated echoes, and the contribution of radio-frequency pulse fronts distribution mechanism is insignificant. For this reason lithium ferrite is a good material for the experimental verification of theoretical conclusions in experimental conditions most close to the theoretical model.

# 1. Introduction

It is known that in the magnetically ordered systems the formation of NMR responses after the excitation of nuclear spin systems (NSS) of a sample by RF pulse sequence is defined by two main reasons: inhomogeneous broadening (IB) of the spectroscopic transitions (so-called Zeeman inhomogeneous broadening) and inhomogeneous distribution of the radio-frequency (RF) field enhancement factor on nuclei (so-called Rabi IB), caused by the hyperfine interaction between electron and nuclear spins [1].

For investigations of the echo-responses of Hahn spin-systems (further we consider conditions where the dynamic frequency shift does not play a role [1]) it was devoted very many works [2-8]. Theoretical and experimental definition of time moments of the echo-signal appearance, their number and intensity makes it possible to identify the echo formation mechanisms in different materials what is important for the practical applications.

# 2. Statement of problem

At starting time moment $t = 0$ the equilibrium NSS of magnet is excited by the first RF pulse of length $\tau_1$. Then during time $\tau > \tau_1$ (delay time) a free evolution of NSS takes place, after which in time moment $t = \tau_1 + \tau$ the second RF pulse of length $\tau_2 > \tau_1 + \tau$ is applied. After termination of the second pulse in the time moment $t = \tau_1 + \tau + \tau_2 = t_1$ the process of free induction starts again and on the time scale $t > t_1$ the multiple echo signals are observed.

It is supposed that pulses have generally different frequencies $\omega^{(1,2)}$, amplitudes $\omega_1^{(1,2)}$ (in the frequency units) and lengths $\tau_{1,2}$. It is understood also that in the time moment of a pulse application it takes place the jumplike change of hyperfine field on nuclei: $H_{nj} \to H_{nj}^{(1,2)}$ (as example, due to the displacement of domain wall as result of magnetic pulse influence [9], or due to other reason), what results in the jumplike change of resonance frequency of the $j$ – isochromat $\omega_j = \gamma_I H_{nj} \to \omega_j^{(1,2)} = \gamma_I H_{nj}^{(1,2)}$ (it is believed that usually $H_{nj} >> H_o$).

In the coordinate system rotating with RF field frequency $\omega$ around $z$ axis (RCS) NSS is described by the density matrix $\rho*(t)$ and Liouville equation [10]

$$i\hbar \frac{\partial \rho*(t)}{\partial t} = \left[ H_t^*, \rho*(t) \right], \qquad (1)$$

where

$$H_t^* = H* + H_{SL}^*(t), \quad H* = \hbar \sum_j \left( \Delta_j I_j^z + \omega_1 I_j^x \right) + H_{SN}, \quad H_{SL}^*(t) = \exp(i\omega I_z t) H_{SL} \exp(-i\omega I^z t); \quad (2)$$

$H_{SL}$ - spin-lattice interaction (we do not need the full form of it), $H_{SN}$ - is the Suhl-Nakamura interaction:

$$H_{SN} = \sum_{k \neq \ell} U_{k\ell} I_k^+ I_\ell^-, \quad U_{k\ell} = U_{\ell k}.$$

At further consideration we neglect the interaction of nuclear spins from different isochromats, therefore we could present $H_{SN}$ in the form:

$$H_{SN} \approx \sum_j (H_{SN})_j, \quad (H_{SN})_j = \sum_{k \neq \ell}^{N_j} U_{k\ell} I_k^+ I_\ell^-, \qquad (3)$$

where the summing over $k, \ell$ is spread on spins of the $j$-th isochromat (their number is $N_j$). It is obvious that

$$\left[ I_j^z, (H_{SN})_j \right] = 0. \qquad (4)$$

Value $\Delta_j = \omega_j - \omega$ is the resonance detuning; $\omega_j = \gamma_n H_j$ is the Zeeman frequency of the $j$-th isochromat; $\omega_1 = \eta \gamma_I H_1$ is the Rabi frequency (the Rabi IB is not taken into account); $\eta$ – the RF field enhancement factor; $I_j^{x,y,z}$, $I_j^\pm = I_j^x \pm i I_j^y$ is nuclear spin operators belong to the $j$-th isochromat.

As it is seen from Hamiltonian $H^*$ in form (2) the nuclear spins in RCS experience influence of the effective field:

$$\vec{H}_j = \left( \Delta_j \vec{k} + \omega_1 \vec{i} \right) / \gamma_I. \qquad (5)$$

The action of two different RF pulses on NSS could be schematically presented in the following way (Fig.1):

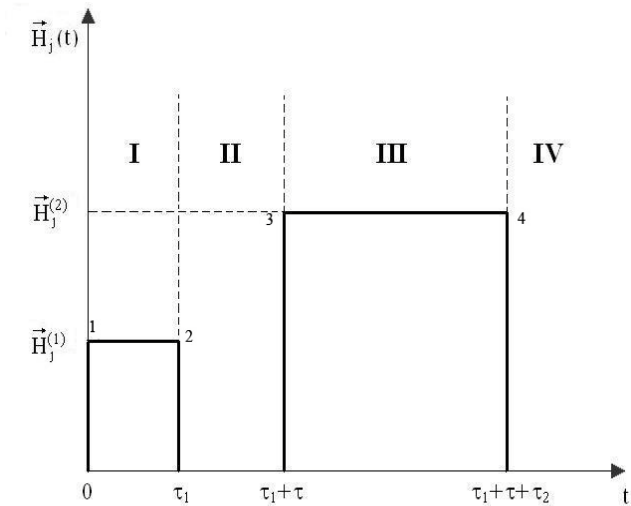

Fig.1. Schematic presentation of the influence of two wide RF pulses on NSS $\tau > \tau_1$, $\tau_2 > \tau_1 + \tau$,
1,2,3,4 - pulse fronts.

The problem is in the solution of the Liouville equation (1) in regions $I - IV$ at the initial condition

$$\rho^*(t)\big|_{t=0} = \rho_0 \cong \left( 1 - \hbar \beta_L \sum_j \omega_j I_j^z \right) / Tr1, \qquad (6)$$

allowing for the condition of concatenation in time moments $\tau_1$, $\tau_1+\tau$, $\tau_1+\tau+\tau_2$:

$$\rho_1^*(0)=\rho_0, \quad \rho_{10}^*(\tau_1)=\rho_1^*(\tau_1), \quad \rho_2^*(\tau_1+\tau)=\rho_{10}^*(\tau_1+\tau), \tag{7}$$

$$\rho_{20}^*(\tau_1+\tau+\tau_2)=\rho_2^*(\tau_1+\tau+\tau_2),$$

providing the continuity of solution.

During the action of RF pulses the spin-lattice interaction $H_{SL}^x(t) \ll H_{SN}$ is neglected.

The NSS Hamiltonians in regions $I(\alpha=1)$ and $III(\alpha=2)$ are presented by expressions.

$$H_{t\alpha}^* = H_\alpha^* + H_{SL,\alpha}^*(t), H_\alpha^* = \hbar \sum_j \left( \Delta_{j\alpha} I_j^z + \omega_1^{(\alpha)} I_j^x \right), \quad \alpha=1, 2. \tag{8}$$

$H_{t\alpha}^*$ is obtained from $H_t^*$ in (2) with help of substitutions

$$\omega \to \omega^{(\alpha)}, \; \omega_1 \to \omega_1^{(\alpha)}, \; \Delta_j \to \Delta_{j\alpha} = \omega_j^{(\alpha)} - \omega^{(\alpha)}, \; \omega_j \to \omega_j^{(\alpha)} = \gamma_I H_{nj}^{(\alpha)} \tag{9}$$

In regions $II$ and $IV$ Hamiltonian $H_0^*(t)$ is obtained from $H_t^*$ at $\omega_1=0$ and takes form:

$$H_0^*(t) = \hbar \sum_j \Delta_j I_j^z + H_{SN} + H_{SL}^*(t) \tag{10}$$

From the form of $H_\alpha^*$ it follows that during the action of $\alpha$–th pulse the nuclear spins are influenced by the effective magnetic field

$$\vec{H}_j^{(\alpha)} = \left( \Delta_{j\alpha} \vec{k} + \omega_1^{(\alpha)} \vec{i} \right) / \gamma_I, \quad \alpha=1, 2, \tag{11}$$

where $\vec{k}, \vec{i}$ are orths of $z$, $x$ exes in the RCS.

### 3. Solution of Liouville equation

### *Region I*

The solution of equation

$$i\hbar \frac{\partial \rho_1^*(t)}{\partial t} = \left[ H_1^*, \rho_1^*(t) \right] \tag{12}$$

at the initial condition (6) takes form:

$$\rho_1^*(t) = \exp\left( -\frac{i}{\hbar} H_1^* t \right) \rho_0 \exp\left( \frac{i}{\hbar} H_1^* t \right) \tag{13}$$

Let us transform expression (13) to a more convenient form for calculations introducing with this aim a unitary operator

$$U_{y1} = \exp\left(i\sum_{j}\theta_{j1}I_j^y\right), \quad \cos\theta_{j1} = \Delta_{j1}/\Omega_{j1}, \quad \Omega_{j1} = \sqrt{\Delta_{j1}^2 + (\omega_1^{(1)})^2} \equiv \gamma_I H_j^{(1)}. \tag{14}$$

The operator $U_{y1}$ realizes a rotation of coordinate system around $y$ axis on $\theta_{j1}$ angles different for different isochromats. At this rotation for each isochromat the $z$ axis of "new" system becomes directed along the effective magnetic field $\vec{H}_{j1}$ from (11).

It is easy to show that

$$H_1^* = U_{y1}^+ \tilde{H}_{z1} U_{y1} + H_{SN} = U_{y1}^+ \{\tilde{H}_{z1} + U_{y1} H_{SN} U_{y1}^+\} U_{y1}, \quad \tilde{H}_{z1} = \hbar \sum_{j}\Omega_{j1} I_j^z. \tag{15}$$

Neglecting in the "modified" interaction $U_{y1} H_{SN} U_{y1}^+$ by the nonsecular, i.e. noncommuting with $\tilde{H}_{j1}$, terms (it is supposed that $\omega_1 \gg \omega_{SN} = \left\{\dfrac{TrH_{SN}^2}{\hbar^2 TrI_z^2}\right\}^{1/2} \sim \dfrac{1}{T_2}$, $T_2^{-1}$ is "homogeneous" NMR spectrum width) we obtain

$$H_1^* = U_{y1}^+ \tilde{H}_1 U_{y1}, \quad \tilde{H}_1 = \tilde{H}_{z1} + \tilde{H}_{SN,1}, \quad H_{SN,1} = \sum_{j}\sum_{k\neq\ell}^{N_j} U_{k\ell}\{\lambda(\theta_{j1}) I_k^+ I_\ell^- + \sin^2\theta_{j1}\vec{I}_k\cdot\vec{I}_\ell\}. \tag{16}$$

Allowing for (13) and (16), we obtain finally

$$\rho_1^*(t) = U_{j1}^+ \exp\left(-\frac{i}{\hbar}\tilde{H}_1 t\right) U_{y1} \rho_0 U_{y1}^+ \exp\left(\frac{i}{\hbar}\tilde{H}_1 t\right) U_{y1}. \tag{17}$$

### *Region II*

It is necessary to solve the equation

$$i\hbar\frac{\partial \rho_{10}^+(t)}{\partial t} = [H_0^*(t), \rho_{10}^*(t)] \tag{18}$$

with Hamiltonian (10) at initial condition

$$\rho_{01}^*(t)\Big|_{t=\tau_1} = \rho_1^*(\tau_1). \tag{19}$$

Let us write Hamiltonian $H_0^*(t)$ in form

$$H_0^*(t) = \hbar\sum_{j}\Delta_j I_j^z + V(t), \quad V(t) = H_{SN} + H_{SL}^*(t),$$

and pass to a new presentation

$$\rho'(t) = exp\left(i\sum_j \Delta_j I_j^z (t-\tau_1)\right) \rho_{10}^*(t) exp\left(-i\sum_j \Delta_j I_j^z (t-\tau_1)\right). \qquad (20)$$

It is evident, that

$$\rho'(t)\big|_{t=\tau_1} = \rho_{10}^*(\tau_1) = \rho_1^*(\tau_1). \qquad (21)$$

For the operator $\rho'(t)$ one obtains equation

$$i\hbar \frac{\partial \rho'(t)}{\partial t} = [V'(t), \rho'(t)], \quad V'(t) = H_{SN} + exp\left(i\sum_j \Delta_j I_j^z (t-\tau_1)\right) H_{SL}^*(t) exp\left(-i\hbar \sum_j \Delta_j I_j^z (t-\tau_1)\right). \qquad (22)$$

Solving equation (22) at initial condition (21) and using (20), we find out

$$\rho_{10}^*(t) = exp\left(-i\sum_j \Delta_j I_j^z (t-\tau_1)\right) U(t,\tau_1) \rho_1^*(\tau_1) U^+(t,\tau_1) exp\left(i\sum_j \Delta_j I_j^z (t-\tau_1)\right)$$

$$U(t,\tau_1) = T\, exp\left(-\frac{i}{\hbar} \int_{\tau_1}^{t} V'(t') dt'\right). \qquad (23)$$

Evidently, the expression $exp\left(-i\sum_j \Delta_j I_j^z (t-\tau_1)\right) U(t,\tau_1)$ is the evolution operator for NSS in the time interval $\tau_1 \le t \le \tau_1 + \tau$ between pulses

### Region III

The solution of equation

$$i\hbar \frac{\partial \rho_2^*(t)}{\partial t} = [H_2^*, \rho_2^*(t)] \qquad (24)$$

at initial condition

$$\rho_2^*(\tau_1 + \tau) = \rho_{10}^*(\tau_1 + \tau), \qquad (25)$$

in analogy with (17) takes the form

$$\rho_2^*(t) = U_{y2}^+ exp\left(-\frac{i}{\hbar} \tilde{H}_2 (t-\tau_1-\tau)\right) U_{y2} \rho_{10}^*(\tau_1+\tau) U_{y2}^+ exp\left(\frac{i}{\hbar} \tilde{H}_2 (t-\tau_1-\tau)\right). \qquad (26)$$

## Region IV

Solution of equation

$$i\hbar \frac{\partial \rho_{20}^*(t)}{\partial t} = [H_0^*(t), \rho_{20}^*(t)] \tag{27}$$

at initial condition

$$\rho_{20}^*(\tau_1 + \tau_2 + \tau) = \rho_2^*(\tau_1 + \tau + \tau_2) \tag{28}$$

in analogy with (23), takes the form

$$\rho_{20}^*(t) = \exp\left(-i\sum_j \Delta_{jj} I_j^z (t - \tau_1 - \tau - \tau_2)\right) U(t, \tau_1 + \tau + \tau_2) \rho_2^*(\tau_1 + \tau + \tau) \cdot$$
$$\cdot U^+(t, \tau_1 + \tau + \tau_2) \exp\left(i\sum_j \Delta_j I_j^z (t - \tau_1 - \tau - \tau_2)\right). \tag{29}$$

**3.**

**4. General expression for free induction decay (FID) signal**

The FID signal is proportional to the value of transverse nuclear magnetization

$$I(t) = Sp\{\rho_{20}^*(t) I^+\} = \sum_j \exp[i\Delta_j (t - \tau_1 - \tau - \tau_2)] \cdot$$
$$\cdot Sp\{\rho_2^*(\tau_1 + \tau + \tau_2) U^+(t, \tau_1 + \tau + \tau_2) I_j^+ U(t, \tau_1 + \tau + \tau_2)\}. \tag{30}$$

For obtaining (30) we use the relation

$$\exp\left\{i\sum_j \Delta_j I_j^z (t - \tau_1 - \tau - \tau_2)\right\} I^+ \exp\left\{-i\sum_j \Delta_j I_j^z (t - \tau_1 - \tau - \tau_2)\right\} = \sum_j \exp\{i\Delta_j (t - \tau_1 - \tau - \tau_2)\} I_j^+$$

As far as $H_{SL} \ll H_{SN}$, one could take in (30)

$$U(t, \tau_1 + \tau + \tau_2) \cong \exp\left\{-\frac{i}{\hbar} H_{SN} (t - \tau_1 - \tau - \tau_2)\right\}. \tag{31}$$

Making in (30) the substitution $t \to t + t_1$, $t_1 = \tau_1 + \tau + \tau_2$ and allowing for (31), we obtain

$$I(t + t_1) = \sum_j \exp(i\Delta_j t) Sp\{\rho_2^*(t_1) I_j^+(t)\}, \quad I_j^+(t) = \exp\left(\frac{i}{\hbar} H_{SN} t\right) I_j^+ \exp\left(-\frac{i}{\hbar} H_{SN} t\right). \tag{32}$$

In formula (32) time $t$ is counted from a time moment of the second pulse termination $t_1$.

Accordingly to (26)

$$\rho_2^*(t) = U_{y2}^+ \exp\left(-\frac{i}{\hbar}\tilde{H}_2\tau_2\right) U_{y2} \rho_{10}^*(\tau_1+\tau) U_{y2}^+ \exp\left(\frac{i}{\hbar}\tilde{H}_2\tau_2\right) U_{y2}. \tag{33}$$

The substitution of (33) to (32) gives:

$$I(t+t_1) = \sum_j \exp(i\Delta_j t) Sp\left\{\rho_{10}^*(\tau_1+\tau) U_{y2}^+ \exp\left(\frac{i}{\hbar}\tilde{H}_2\tau_2\right) U_{y2} I_j^+(t) U_{y2}^+ \exp\left(-\frac{i}{\hbar}\tilde{H}_2\tau_2\right)\right\}. \tag{34}$$

A further simplification of expression (34) is made by the approximation

$$I_j^+(t) = \exp(-\lfloor t \rfloor/T_2) I_j^+, \tag{35}$$

which is equivalent to the Lorentz approximation of the correlation,

$$Sp\{I^- I^+(t)\}/Sp1 = Sp\{I^- I^+\}/Sp1 \cdot \exp(-\lfloor t \rfloor/T_2),$$

frequently used in practice [11]. In this approximation the transverse magnetization in region *III* coincides with the solution of Bloch equations for this region. As a result, one obtains:

$$I(t+t_1) = \sum_j \exp\left[\left(i\Delta_j - \frac{1}{T_2}\right)t\right] Sp\left\{\rho_{10}^*(\tau_1+\tau) U_{y2}^+ \exp\left(\frac{i}{\hbar}\tilde{H}_2\tau_2\right) \cdot U_{y2} I_j^+ U_{y2}^+ \exp\left(-\frac{i}{\hbar}\tilde{H}_2\tau_2\right) U_{y2}\right\}. \tag{36}$$

It was shown in work [10] that

$$U_{y2}^+ \exp\left(\frac{i}{\hbar}\tilde{H}_2\tau_2\right) U_{y2} I_j^+ U_{y2}^+ \exp\left(-\frac{i}{\hbar}\tilde{H}_2\tau_2\right) U_{y2} = \exp\left(\frac{i}{\hbar}H_2^*\tau_2\right) I_j^+ \exp\left(-\frac{i}{\hbar}H_2^*\tau_2\right) =$$
$$= \alpha_j^{(2)} I_j^+ + \beta_j^{(2)} I_j^- + \gamma_j^{(2)} I_j^z. \tag{37}$$

Allowing for (37) the expression (36) is transformed to the form:

$$I(t+t_1) = \sum_j \exp\left[\left(i\Delta_j - \frac{1}{T_2}\right)t\right] Sp\{\rho_{10}^*(\tau_1+\tau)[\alpha_j^{(2)} I_j^+ + \beta_j^{(2)} I_j^- + \gamma_j^{(2)} I_j^z]\}. \tag{38}$$

Accordingly to (23),

$$\rho_{10}^*(\tau_1+\tau) = \exp\left(-i\sum_j \Delta_j I_j^z \tau\right) U(\tau_1+\tau,\tau_1) \rho_1^*(\tau_1) U^+(\tau_1+\tau,\tau_1) \exp\left(i\sum_j \Delta_j I_j^z \tau\right). \tag{39}$$

Substituting (39) into (38) and allowing for the relation

$$\exp\left(i\sum_{j'} \Delta_{j'} I_{j'}^z \tau\right) I_j^\pm \exp\left(-i\sum_{j'} \Delta_{j'} I_{j'}^z \tau\right) = \exp(\pm i\Delta_j \tau) I_j^\pm,$$

we obtains

$$I(t+t_1) = \sum_j \exp\left((i\Delta_j - T_2^{-1})t\right) Sp\{\rho_1^*(\tau_1) U^+(\tau_1+\tau,\tau_1)[\alpha_j^{(2)} I_j^+ \exp(i\Delta_j\tau) + \beta_j^{(2)} I_j^- \exp(-i\Delta_j\tau) +$$

$$+\gamma_j^{(2)}I_j^z\bigl]U(\tau_1+\tau,\tau_1)\bigr\}. \qquad (40)$$

Let us use approximations

$$U^+(\tau_1+\tau,\tau_1)I_j^\pm U(\tau_1+\tau,\tau_1)\cong I_j^\pm \exp(-|\tau|/T_2),$$

$$U^+[\tau_1+\tau,\tau_1]I_j^z U(\tau_1+\tau,\tau_1)\cong \bar{I}_j^z+\bigl(I_j^z-\bar{I}_j^z\bigr)\exp(-|\tau|/T_1^{-1}). \qquad (41)$$

The first one of them is equivalent to (35), the second one provides coincidence of the longitudinal magnetization with the solution of Bloch equations in the region $II$. The value $\bar{I}_{j0}^z$ is an equilibrium value of the longitudinal component of nuclear spin in the high-temperature approximation:

$$\bar{I}_j^z = Sp(\rho_0 I_j^z) = -\frac{1}{3}\hbar\beta_L\omega_j S(S+1). \qquad (42)$$

As a result, for the FID signal (40) we obtain the value

$$I(t+t_1)=\sum_j \exp\bigl((i\Delta_j-T_2^{-1})t\bigr)Sp\bigl\{\rho_1^*(\tau_1)\bigl[\alpha_j^{(2)}I_j^+\exp\bigl((i\Delta_j-T_2^{-1})\tau\bigr)+\beta_j^{(2)}I_j^-\exp\bigl((-i\Delta_j-T_2^{-1})\tau\bigr)I_j^- +$$

$$+\gamma_j^{(2)}\bar{I}_j^z+\gamma_j^{(2)}\bigl[(I_j^z-\bar{I}_j^z)\exp(-\tau/T_1)\bigr]\bigr\}. \qquad (43)$$

Substituting in (43) the value $\rho_1^*(\tau_1)$ from (17) and carrying out calculations, we obtain finally the FID signal in the general form:

$$I(t+t_1)=\sum_j \bar{I}_j^z\exp\bigl[(i\Delta_j-T_2^{-1})t\bigr]\bigl\{\gamma_j^{(2)}(1-\exp(-\tau/T_1))+\alpha_j^{(2)}\gamma_j^{(1)*}\exp\bigl[(i\Delta_j-T_2^{-1})\tau\bigr]+$$

$$+\beta_j^{(2)}\gamma_j^{(1)*}\exp\bigl[(-i\Delta_j-T_2^{-1})\tau\bigr]+\gamma_j^{(2)}\bar{\gamma}_j^{(1)}\exp(-\tau/T_1)\bigr\}. \qquad (44)$$

At $\tau=0$ it is obtained the value

$$I(t+\tau_1+\tau_2)=\sum_j \bar{I}_j^z\exp\bigl[(i\Delta_j-T_2^{-1})t\bigr]\bigl\{\alpha_j^{(2)}\gamma_j^{(1)}+\beta_j^{(2)}\gamma_j^{(1)*}+\gamma_j^{(2)}\bar{\gamma}_j^{(1)}\bigr\},$$

which exactly coincides with formula (21) from work [10], as one should expect. Similar to that as the situation with one stepwise change of effective field $\vec{H}_j$ [10] in time moment $t=\tau_1$ after the switching on a RF pulse with duration $\tau=\tau_1+\tau_2$ at moment $t=0$, turned out to be analogous to the excitation of NSS by three short-time pulses [12] (three signals of two-pulse echo (TPE) (12), (13), (23) and one signal of stimulated echo (123) – designations are taken from work [12]), the considered in this work case of two arbitrary pulses appeared to be analogous to the excitation of NSS by four short-time pulses [13]. Substituting values of factors $\alpha_j^{(2)}$, $\beta_j^{(2)}$, $\gamma_j^{(2)}$, $\gamma_j^{(1)}$, $\bar{\gamma}_j^{(1)}$ from work [10] in (44), one could find out that expression for $I(t+t_1)$ contains 20 signals: two induction signals, corresponding time moments $t=\tau_1$, $t=t_1=\tau_1+\tau+\tau_2$ and 18 echo signals, the timing of which is presented in the table:

**Table.** The timing of echo signals arrangement

$t_e$ - the appearance times of echo in case of presence of sudden jumps; $t_e(\Omega_{ja} = \Omega_j)$ - echo appearance times in absence of jumps; $t_e(\Omega_{ja} = \Omega_j; \omega_1 = 0)$ - echo appearance times in case of four short RF pulses [13] (in our designations).

| | | $t_e$ | $t_e(\Omega_{ja} = \Omega_j)$ | $t_e(\Omega_{ja} = \Omega_j; \omega_1 = 0)$ |
|---|---|---|---|---|
| 1 | (12) | $(\Omega_{j1}\tau_1 - \Omega_{j2}\tau_2)/\Delta_j - \tau$ | $\Omega_j(\tau_1 - \tau_2)/\Delta_j - \tau$ | $\tau_1 - \tau_2 - \tau$ |
| 2 | ((12)3) | $\tau - (\Omega_{j1}\tau_1 + \Omega_{j2}\tau_2)/\Delta_j$ | $\tau - \Omega_j(\tau_1 + \tau_2)/\Delta_j$ | $\tau - \tau_1 - \tau_2$ |
| 3 | (123) | $(\Omega_{j1}\tau_1 - \Omega_{j2}\tau_2)/\Delta_j$ | $\Omega_j(\tau_1 - \tau_2)/\Delta_j$ | $\tau_1 - \tau_2$ |
| 4 | (23) | $\tau - \Omega_{j2}\tau_2/\Delta_j$ | $\tau - \Omega_j\tau_2/\Delta_j$ | $\tau - \tau_2$ |
| 5 | (13) | $\tau + (\Omega_{j1}\tau_1 - \Omega_{j2}\tau_2)/\Delta_j$ | $\tau + \Omega_j(\tau_1 - \tau_2)/\Delta_j$ | $\tau + \tau_1 - \tau_2$ |
| 6 | ((12)34) | $\tau - \Omega_{j1}\tau_1/\Delta_j$ | $\tau - \Omega_j\tau_1/\Delta_j$ | $\tau - \tau_1$ |
| 7 | (124) | $\Omega_{j1}\tau_1/\Delta_j$ | $\Omega_j\tau_1/\Delta_j$ | $\tau_1$ |
| 8 | ((13)4) | $(\Omega_{j2}\tau_2 - \Omega_{j1}\tau_1)/\Delta_j - \tau$ | $\Omega_j(\tau_2 - \tau_1)/\Delta_j - \tau$ | $\tau_2 - \tau_1 - \tau$ |
| 9 | (234) | $\tau$ | $\tau$ | $\tau$ |
| 10 | ((23)4) | $\Omega_{j2}\tau_{21}/\Delta_j - \tau$ | $\Omega_j\tau_2/\Delta_j - \tau$ | $\tau_2 - \tau$ |
| 11 | (134) | $\Omega_{j1}\tau_1/\Delta_j + \tau$ | $\Omega_j\tau_1/\Delta_j + \tau$ | $\tau_1 + \tau$ |
| 12 | ((123)4) | $(\Omega_{j2}\tau_2 - \Omega_{j1}\tau_1)/\Delta_j$ | $\Omega_j(\tau_2 - \tau_1)/\Delta_j$ | $\tau_2 - \tau_1$ |
| 13 | (((12)3)4) | $(\Omega_{j2}\tau_2 + \Omega_{j1}\tau_1)/\Delta_j - \tau$ | $\Omega_j(\tau_1 + \tau_2)/\Delta_j - \tau$ | $\tau_1 + \tau_2 - \tau$ |
| 14 | (34) | $\Omega_{j2}\tau_2/\Delta_j$ | $\Omega_j\tau_2/\Delta_j$ | $\tau_2$ |
| 15 | ((12)4) | $(\Omega_{j2}\tau_2 - \Omega_{j1}\tau_1)/\Delta_j + \tau$ | $\Omega_j(\tau_2 - \tau_1)/\Delta_j + \tau$ | $\tau_2 - \tau_1 + \tau$ |
| 16 | (1234) | $(\Omega_{j2}\tau_2 + \Omega_{j1}\tau_1)/\Delta_j$ | $\Omega_j(\tau_1 + \tau_2)/\Delta_j$ | $\tau_1 + \tau_2$ |
| 17 | (24) | $\Omega_{j2}\tau_2/\Delta_j + \tau$ | $\Omega_j\tau_2/\Delta_j + \tau$ | $\tau_2 + \tau$ |
| 18 | (14) | $(\Omega_{j2}\tau_2 + \Omega_{j1}\tau_1)/\Delta_j + \tau$ | $\Omega_j(\tau_1 + \tau_2)/\Delta_j + \tau$ | $\tau_1 + \tau_2 + \tau$ |

The echo signals could be classified in the following way [13]:

1) six primary signals TPE (12), (13), (14), (23), (24), (34), formed in pairs by fronts 1, 2, 3, 4 (see Fig.1). ((12) and (34) are single-pulse echo (SPE) signals from the first and the second RF pulses, correspondingly);

2) four secondary signals of TPE ((12)3), ((12)4), ((13)4), ((23)4), formed by the primary echo signal with following fronts;

3) four signals of primary stimulated echo (123), (124), (134), (234), formed by three fronts;

4) two signals of the secondary TPE (((12)3)4), ((123)4), formed by echo signal ((12)3) and stimulated echo signal (123) with front 4, respectively;

5) one signal of "complicated" stimulated echo ((12)34), formed by three influences: (12) and 3 and 4 fronts;

6) one signal (1234) which is specific for four-pulse sequence and does not formed by one of the above noted mechanisms.

From the presented table it follows that at fulfilling conditions

$$\tau > \tau_1, \quad \tau_2 > \tau_1 + \tau \tag{45}$$

the maximum number of echo signals which could be usually observed after a time moment of the second pulse termination, as in works [6,13], is equal to thirteen (it is not observed the first echo signals (12), ((12)3), (123), (23), (13), the appearance time moments of which are arranged from left side in respect to the moment $t_1 = \tau_1 + \tau + \tau_2$). But the real number of the observed signals and time moments of their appearances depends on relation between $\tau_1, \tau, \tau_2$ and values of jumps.

If instead of (45) one takes condition

$$\tau_1 > \tau + \tau_2, \quad \tau < \tau_2,$$

then the maximum number of the observed echo signals is twelve (it is not observed signals (23), ((12)34), ((12)3), ((12)4), ((123)4), (((12)3)4). But the real number of the observed echo signals and time moments of their appearances depend on relation between $\tau_1, \tau, \tau_2$ and values of jumps. So, as example, at conditionally chosen by us values $\tau_1 = 2, \ \tau = \tau, \ \tau_2 = 7$ for four short-time similar pulses [13] signals (124) and ((13)4) and (134) and ((123)4) appears in pairs in the same place, therefore, it is observed 11, but not 13 echo signals.

So, one could control the number of echo signals and their arrangement changing $\tau_1, \tau, \tau_2, \omega_1$. It should be particularly noted that these changes do not influence the position of stimulated echo (234) ($t_e = \tau$). At $\tau = 0$ one observes four echo signals from work [10].

For shortness, we do not present here all expressions for thirteen echo signals but restrict ourselves with four most interesting echo amplitudes (numbering of echo amplitudes corresponds to their position in the table).

$$((12)34) \to A_{6,e} = \frac{1}{2}\sum_j \bar{I}_j^z \sin\theta_{j1} \sin^2\theta_{j2} \sin^2\theta_{j2}/2 \cdot \exp\left(-\frac{t+\tau}{T_2} - \frac{\tau_1}{T_{2j}^{(1)}}\right).$$

In conditions of (45) the signal of "complicated" stimulated echo ((12)34) is interesting due to the fact that from it there starts the spectrum of echo signals and, besides this, at sufficiently large values $\omega_1^{(1)}$, for which $t_e - \tau - \Omega_{j1}\tau_1/\Delta_j \leq 0$, it is not observed. This signal could be used for evaluation of $\omega_1$ values.

$$(234) \to A_{9,e} = \frac{1}{4}\sum_j \bar{I}_j^z \sin 2\theta_{j1} \sin^2\theta_{j1} \cdot \exp\left(-\frac{t+\tau}{T_2}\right).$$

The primary stimulated echo (234) is notable by fact that the time moment of its appearance $t_e=\tau$ depends only on the delay time, but its amplitude does not depend on $T_1$

$$(1234) \to A_{16,e} = \frac{1}{2}\sum_j \bar{I}_j^z \sin^2\theta_{j1} \sin\theta_{j2} \sin^2\theta_{j2}/2 \cdot \exp\left(-\frac{t}{T_2} - \frac{\tau_1}{T_{2j}^{(1)}} - \frac{\tau_2}{T_{2j}^{(2)}} - \frac{\tau}{T_1}\right).$$

The echo signal (1234) could be used for measurements of spin-relaxation time $T_1$

$$(14) \to A_{18,e} = \sum_j \bar{I}_j^z \sin\theta_{j2} \sin^2\theta_{j2}\left[1 - \exp(-\tau/\pi) \cdot \sin^2\theta_{j1}\right] \cdot \exp\left(-t/T_2 - \tau_2/T_{2j}^{(2)}\right).$$

The primary (TPE) (14) is the last observed echo signal at arbitrary times $\tau_1, \tau, \tau_2$ and jumps.

It should be noted that from thirteen observed echo signals only four ((124), ((123)4), (34) and (1234), depend on $T_1$ which is the consequence of the duration of RF pulses and condition $H_{SL} \ll H_{SN}$.

### 5. Experimental results

Let us present now the experimental results of investigation of multiple echo in lithium ferrite and cobalt formed at excitation by two wide arbitrary width RF pulses. Experiments were carried out using the nuclear spin echo spectrometer and sample of lithium ferrite, enriched by $^{57}$Fe isotope, and cobalt described in [10]. Nuclear spin echo signals were averaged by a "Tektronix 2430 A".

The choice of sample is caused by the different role of pulse fronts in these materials stipulating the existence of two different mechanisms of SPE formation – multiple – pulse and distortion ones [14].

We use approach developed in work [10] where it was investigated multiple-pulse analogs of SPE obtained at jumplike changes o frequency and amplitude of RF pulses, in the limits of the RF pulse length. It was turned out that RF pulse fronts and locations of jumplike changes of $H_{eff}$ in RCS in limits of RF pulse action have their qualitative analogs in the exciting RF pulses of the Hahn echo method. In this approach the value of $H_{eff}$ direction change in RCS is an analog of deflection angle of the nuclear magnetization vector under the influence of two RF pulses in Hanh method. In frames of this approach, the excitation by two wide RF pulses could be considered to be equivalent to the excitation by a complicated single-pulse when besides fronts of RF pulse in limits of its length there are two jumplike changes of $H_{eff}$ in RCS, with the amplitude of RF pulse between them being zero. This complicated single-pulse excitation has its analog in the four-pulse excitation in the Hahn method [13,15].

Oscillograms of multiple echo signals at excitation by two wide RF pulses in lithium ferrite are presented in Fig.2.

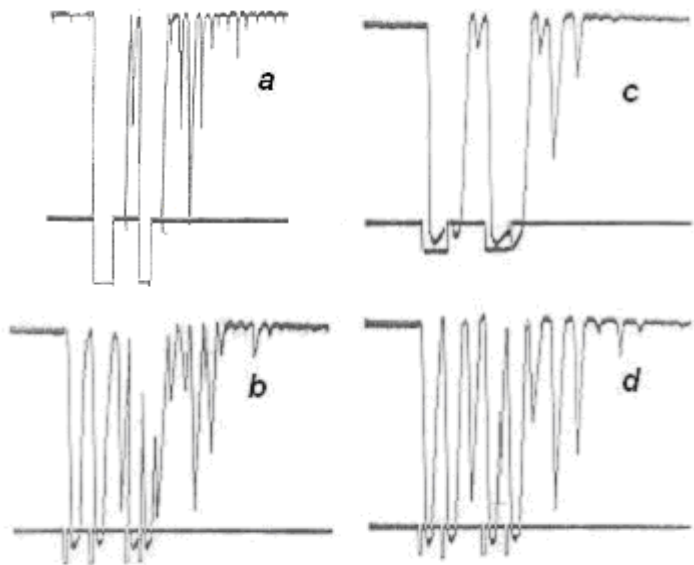

Fig.2. a) at $f_{NM}$ = 71 MHz. RF pulse durations $\tau_1$=8 μs, $\tau_2$=5 μs and time interval between them $\tau_{12}$=9 μs.

```
b) at excitation by four short τp=1 μs RF pulses coinciding with the edges of
wide pulses. c), d) correspond to the case of two equal-length RF pulses.
```

The repetition period of RF pulse pairs is optimal for observation of multiple echo. The upper beam shows echo signal amplitudes from NMR receiver in dependence of time and lower beam – signals from video-detector showing the shape, amplitude and duration of RF pulses.

In Fig.2a it is presented the oscillograms of multiple echo signal obtained at excitation by two RF pulses of different durations and in Fig.2b – the oscillograms of its four-pulse analog in case when four short RF pulses coincide with fronts of two wide RF pulses.

In Fig.2c and 2d it is presented corresponding oscillograms of multiple echo and its four-pulse analogs in lithium ferrite for two equal length RF pulses. In this case the picture is essentially simplified and similar the one considered in [10].

Similar oscillograms for cobalt are presented in Fig.3 a,b.

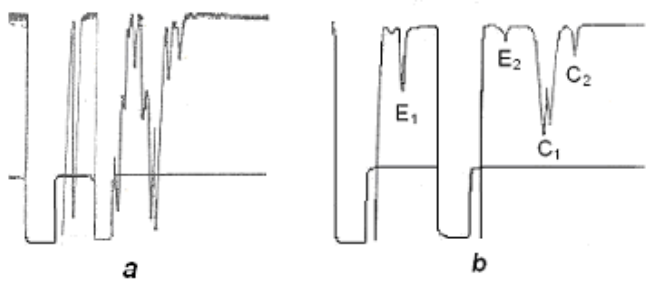

Fig.3. Multiple echo signals in cobalt at excitation by two wide RF pulses with frequency $f_{NMR}$ = 217 MHz
(a) $\tau_1$=10 μs, $\tau_2$=7 μs, $\tau_{12}$=16 μs; (b) $\tau_1$ = $\tau_2$ = $\tau_p$ =10 μs, $\tau_{12}$=16 μs.

The multiple echo signals oscillograms in lithium ferrite [6], for multiple echo signals in $^{59}$CoCoFeNi and $^{51}$VFe obtained on a wide-band coherent NMR spectrometer in the mode of phase-sensitive detection at 4.2K, Fig.4. The multiple echoes in both cases have thirteen components.

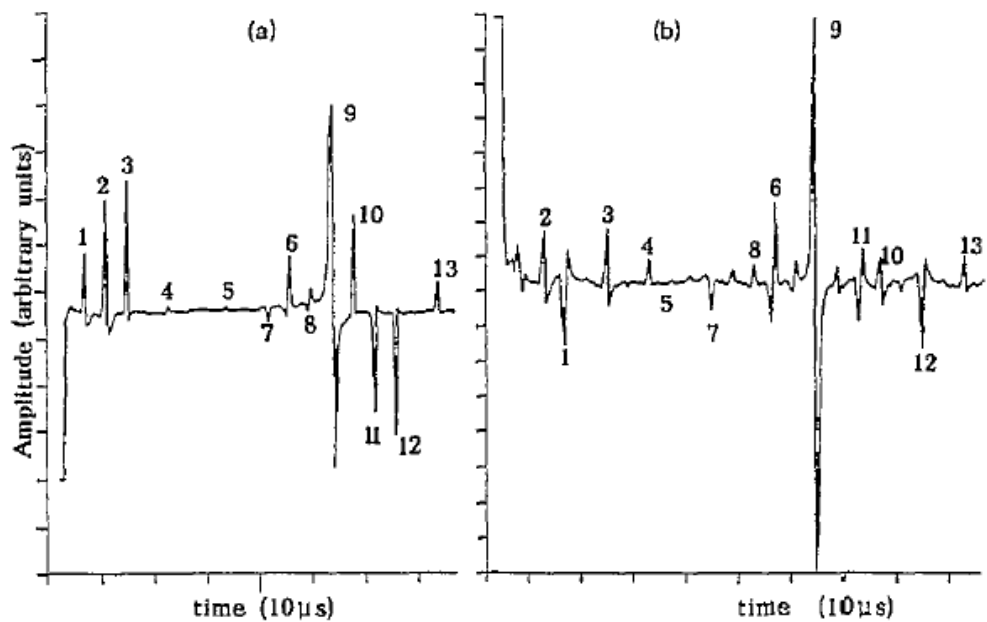

Fig.4. Signals of multiple echo (phase-sensitive detection at 4.2 K) in (a) $^{59}$CoCoFeNi on 220 MHz frequency, repetition rate 10 msec, $\tau_1$=12 μs, $\tau_2$=8 μs, $\tau$=51 μs.(b) $^{51}$VFe, frequency 98 MHz, repetition period 5 msec, $\tau_1$=20 μs, $\tau_2$=8 μs, $\tau$=74 μs [6].

At the same time the multiple echo components in cobalt differ significantly from ones in lithium ferrite by their significantly longer widths, and its main component coinciding with TPE signal in the limit of short RF pulses has the shape of two-hump Mims echo signal [2]. In difference with this in lithium ferrite it is observed considerably narrow echo signals with their shape and intensity depending on the repetition frequency of pairs of RF pulses similarly to those observed in [14] for signals of SPE formed by the multiple pulse mechanism.

Even more large difference is revealed at analysis of relaxation characteristics of multiple echo components in lithium ferrite and cobalt. It is simpler to compare them in case of two equal length wide RF pulses. This situation was for the first time considered in [3] on example of FeV. In this work it was observed unusually fast relaxation rate for one of main components of multiple echo ($C_2$ in designations of [3]). The room temperature measurements gave rates of 1, $1 \cdot 10^{-2}$ and $0.3 \cdot 10^{-3}$ μs$^{-1}$, correspondingly, for the three processes: $C_2$ decay, spin-spin and spin-lattice relaxation.

Similar study carried out in work [8] for lithium ferrite gave results close to [3] and made it possible to clear out the nature of strongly relaxing along with RF pulse duration increase component which, as it was turned out, was changing synchronously with single-pulse and two-pulse stimulated echo signals. As compared with work [3], they have larger intensities providing the possibility to carry out relaxation measurements. It was established that this component possessed relaxation rate close to ones of SPE and stimulated TPE which are formed in lithium ferrite by the multiple-pulse mechanism [8]. In the same time, the relaxation rate of corresponding component in Co was close to the one of SPE in cobalt which in correspondence with the distortion mechanism effective in this material has value close to $T_2$ ($0.5 \div 0.8$ $T_2$ [14]).

On the basis of presented results it is possible therefore to carry out the conclusion on similarity of multiple echo formation mechanisms in lithium ferrite and in materials studied in [6].

The presence of so short relaxation times of one of the main component of multiple echo at increase of RF pulse duration could be understood taking into account the fact, firstly noted in [16], that the SPE signal in frames of the nonresonant formation mechanism of SPE could have more short relaxation times then $T_2$ because the dephasing conditions of isochromats in the effective RF field at the action of RF pulse differ from ones during the process of isochromat rephasing after the termination of RF pulse.

## 6. Conclusion

The method of quantum-mechanical calculations developed earlier by us to study the multiple-pulse analog of single-pulse echo in magnets upon sudden jumps of the effective magnetic fields in rotating coordinate system during the action of RF pulse was used for calculations of intensities and time moments of appearance of multiple echo signals at excitation the nuclear spin system of magnets by two arbitrary width RF pulses.

The formation mechanisms of multiple echo signals and influence of relaxation processes are discussed. The total number of echo signals predicted theoretically and confirmed experimentally is thirteen.

Spin echo signals formed at excitation by four short RF pulses coinciding with fronts of two wide RF pulses were generated as analogs of multiple echo signals. Experiments were carried out in lithium ferrite where the RF pulse front distortions are insignificant and compared with ones in cobalt where the contribution of distortion mechanism in our experimental conditions is significant. For this reason lithium ferrite is appropriate material for the experimental verification of theoretical results in experimental conditions most close to the theoretical model.

## Acknowledgments

The work is supported by the Georgian National Science Foundation N GNSF/ST07/7-248 Grant.